\documentclass[
]{ceurart}

\begin{document}

\copyrightyear{2022}
\copyrightclause{Copyright for this paper by its authors.
  Use permitted under Creative Commons License Attribution 4.0
  International (CC BY 4.0).}

\conference{TPDL2022: 26th International Conference on Theory and Practice of Digital Libraries, 20-23 September 2022, Padua, Italy}

\title{A Continual Relation Extraction Approach for Knowledge Graph Completeness}
\author{Sefika Efeoglu}[%
orcid=0000-0002-9232-4840,
email=sefika.efeoglu@fu-berlin.de,
]
\address{Free University of Berlin, Takustrasse 9, 14195 Berlin, Germany}

\cortext[1]{This doctoral thesis is supervised by Prof. Dr. Adrian Paschke at the Free University of Berlin.}

\begin{abstract}
Representing unstructured data in a structured form is most significant for information system management to analyze and interpret it. To do this, the unstructured data might be converted into Knowledge Graphs, by leveraging an information extraction pipeline whose main tasks are named entity recognition and relation extraction. This thesis aims to develop a novel continual relation extraction method to identify relations (interconnections) between entities in a data stream coming from the real world. Domain-specific data of this thesis is corona news from German and Austrian newspapers.
\end{abstract}

\maketitle
\section{Introduction}
For various reasons, semantic analysis and interpretation of unstructured data, such as social media posts, texts on web pages, and chats, are one of the challenges in information management systems. To analyze and interpret unstructured text data, it must be represented in a structured form. One way of representing unstructured data in the structured form is the use of knowledge graphs (KGs)~\cite{Amit_2019}. An information extraction (IE) pipeline is generally organized as the product of several analysis components, e.g.: named entity (NE) tagging; syntactic analysis; coreference resolution within a document; entity, relation and event extraction (semantic analysis); and cross-document coreference resolution~\cite{Grishman_2015}. Semantic analysis in IE systems might be carried out on the KG.

Recent KG construction approaches utilize machine learning-based approaches instead of rule-based techniques for NE and relation extraction (RE). The machine learning-based approaches obtain impressive results in the NE tagging~\cite{Jing_2018}, while they do not show this performance on RE~\cite{AlAswadi2019AutomaticOC}. Detection of relations between entity pairs has been addressed with various types of approaches: (i) supervised techniques including features-based and kernel-based methods, (ii) a special class of techniques which jointly extract entities and relations ( semi-supervised), (iii) unsupervised, (iv) Open IE and (v) distant supervision based techniques~\cite{Sachin_2017}. Supervised techniques require a large annotated data set, and its annotation process is time-consuming and expensive~\cite{Sachin_2017}. Distant supervision is amongst one of the popular methods to deal with this annotated data problem. The distant supervision, based on existing knowledge bases, brings its own drawback, and it faces the problem of wrongly labeled sentences troubling the training due to the excessive amount of noise~\cite{Aydar_2020}. In addition to distant supervision, another most popular approach is bootstrapping-based semantic RE techniques, namely weakly supervised RE~\cite{Eugene_2000}. However, the weakly supervised approach is more error-prone because of semantic drift in a set of patterns per iteration of a snowball algorithm~\cite{Eugene_2000}. On the other hand, in rule-based RE approaches, finding relations are mostly dependent on predefined rules~\cite{Sachin_2017}. 
 
Recently, extracting information from health data published on the web has become more significant to analyze and take precautions due to the ongoing pandemic. To analyze and evaluate the processes of the pandemic or its measurements, its data must be represented in a machine-readable and understandable format. Moreover, the IE pipeline extracted information from web documents or pages must run throughout producing new data by web applications without any interruption.
 
Concerning the applicability of existing RE methods, information systems have to cope with different challenges, leveraging KG in real-world applications in terms of RE: (i) missing extracted relations between the entities throughout the data stream, since the existing methods run once on a fixed data set, (ii) heterogeneous representation of data, (iii) requirement of a large annotated data set and (iv) unexplainable relations extracted by machine learning-based approaches. The existing RE approaches trained and evaluated on the fixed data set are mostly dependent on predefined relations~\cite{Sachin_2017,Aydar_2020,Ji_2022}; therefore, they might not discover new relation types in the applications whose data is coming from the real world. Because of this reason, to discover new relation types, the learning process must be continuous. Otherwise, using the existing RE approaches on real-world applications leads to KG incompleteness~\cite{Ji_2022}.
 
The rest of this paper explains the thesis' problem details in the Problem Statement section, seeking the solution to research questions in this thesis. After that, this paper introduces the possible methods for these research questions in the Research Methodology section, and then the paper gives possible evaluation approaches in the Evaluation section. Lastly, it details concluding remarks in the Conclusion section.

\section{Problem Statement}
As stated in the Introduction section, relation extraction is one of the most significant tasks in knowledge acquisition which is used for managing and analyzing data on web documents and pages in the health domain. Traditional machine learning-based relation extraction approaches such as distant supervision and bootstrapping methods run once on a fixed data set offline. The main problem with these approaches is that it could not be feasible to detect relations between entities throughout a data stream coming from the real world since they run once as offline on the fixed data set~\cite{PARISI201954}. Because of this reason, they might not keep and transfer knowledge learned from previous tasks to further tasks.

Furthermore, existing RE approaches for streaming data like meta-continual learning are applied offline and do not support knowledge retention~\cite{wu2021curriculummeta}. To provide learning over data coming from real-world applications, knowledge retention is one of the most important requirements for human-like learning. In addition, this thesis addresses KG incompleteness for RE in terms of non-stationary text data. This thesis tries to tackle the problem above, seeking answers to the following research questions (RQ):
\begin{itemize}
    \item \textbf{RQ1:} How can new relation types between entities be continuously discovered throughout a data stream coming from the real world? (The ${1}^{st}$)
    \item \textbf{RQ2:} How can the KG incompleteness in terms of relations between entities be addressed in the context of continuous streaming data? (The ${2}^{nd}$)
    \item \textbf{RQ3:} Can semantic drift of a set of patterns in a weakly supervised approach be tackled by a rule learning method using KG or ontology embeddings? (The ${3}^{rd}$)
    \item \textbf{RQ4:} How can finally extracted relations be made explainable and interpretable? since machine learning approaches are black-box. (Last year)
\end{itemize}

\section{Research Methodology}
The thesis proposal hypothesizes that a weakly supervised continual RE approach tackles the KG incompleteness problem in terms of relations between entities, identifying relation types continuously in a data stream coming from real-world applications like newspapers. In addition, this thesis claims that continual learning might increase the number of newly discovered different relation types, feeding the model with new data resources. The proposed research methods below (See Fig. 1) are developed and evaluated on the corona news published on Tagesschau~\footnote{Tagesschau : \url{https://www.tagesschau.de/}} and Austrian Derstandard~\footnote{Derstandard: \url{https://www.derstandard.at/}}.

\begin{figure}[!h]
    \centering
    \includegraphics[width=0.6\textwidth]{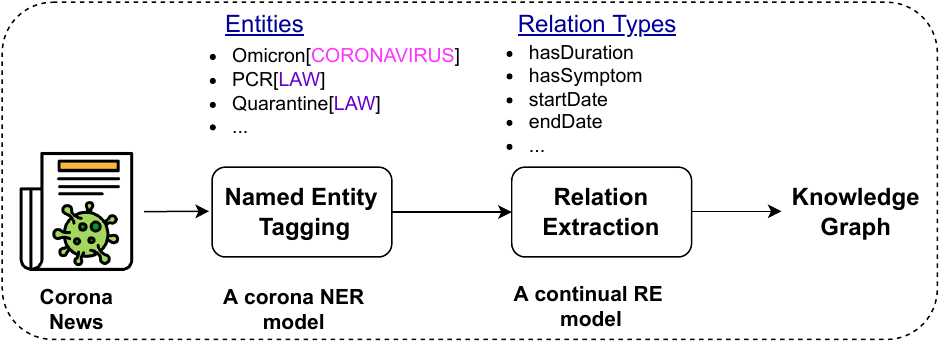}
    \caption{This pipeline shows how the thesis is progressing.}
\end{figure}
\subsection{Named Entity Tagging for Corona News}
Since a corona Named Entity Recognition (NER) model has not been developed yet, and there is no gold standard benchmark data set in this domain, the NER model must be developed to identify named entities in a text before developing and running a RE approach for this data set. However, a weakly supervised CORD-NER data set was published in~\cite{CORD_NER} and is used in the baseline NER model of this thesis after reducing the noise in its labels with Wikidata~\footnote{Wikidata: \url{https://www.wikidata.org}}. Then, in order to improve its accuracy on the corona news text data, fine-tuning is applied to this baseline model with Corona News corpus (their English versions) described above.

\subsection{A Continual Relation Extraction Approach for Knowledge Graph Completeness}
Continuously extraction of relations from non-stationary data still has to address some challenges, such as labeled training data, catastrophic forgetting (because of continual learning) , and predefined relation types. This thesis aims to develop a (online) weakly supervised continual RE approach by using a snowball algorithm which applies an incremental learning method~\cite{Eugene_2000}. However, this algorithm has its own drawbacks like a semantic drift on pattern extraction for relation types, and seed relation selection in the first step of the algorithm. Although there are previous attempts to implement this algorithm for RE, they have fixed constant coefficients in its learning phase or knowledge retention problems due to their rule-based approach~\cite{batista-etal-2015-semi}, and also simple transfer learning~\cite{Tianyu_2019}. Nevertheless, both approaches are evaluated on stationary data for predefined relation types. With respect to evaluation of the continual RE algorithms on the streaming data, existing (offline) continual learning RE algorithms~\cite{wu2021curriculummeta, biesialska-etal-2020-continual} have been tested for the predefined relations on the benchmarks, like FewRel~\cite{han-etal-2018-fewrel} and SimpleQuestions~\cite{petrochuk-zettlemoyer-2018-simplequestions}. As stated earlier, the previous algorithms have been run offline and not supported the knowledge retention. An online continual RE algorithm, e.g., neurogenesis and memory replay might transfer the learnt knowledge (relation types) to next tasks~\cite{PARISI201954}. Therefore, this thesis aims to transfer the learnt knowledge in the previous tasks to the following task with an assist of the continual learning algorithm to keep the learnt relation types. In addition to this, the algorithm will consider knowledge graph and category embeddings to discover new relation types together with dependency parsing of a sentence.

\section{Evaluation Approaches}
This thesis evaluates the continual RE approach's results with evaluation metrics in terms of various perspectives. The thesis takes into account the following metrics: Precision (P), Recall (R), F1 score, P-R curve, area under the curve~\cite{biesialska-etal-2020-continual} to evaluate the relation extraction approach. The existing relation extraction algorithms using curriculum meta continual learning~\cite{wu2021curriculummeta, biesialska-etal-2020-continual} have been evaluated offline on the partitions of benchmark data sets e.g., FewRel~\cite{han-etal-2018-fewrel} and SimpleQuestions~\cite{petrochuk-zettlemoyer-2018-simplequestions} for time slots. However, there is no approach to evaluate online RE algorithms on streaming data. Similarly, this thesis will evaluate the metrics above for the time slots of the data and measure its performances on the benchmark data in the first phase of its implementation. Furthermore, this thesis also considers following metrics to evaluate the approach's performance with respect to continual learning: average accuracy, whole accuracy, forgetting measure, learning curve area, and error bound~\cite{wu2021curriculummeta,biesialska-etal-2020-continual}.

\section{Conclusion}
Consequently, this thesis is planning to develop a continual learning approach to identify relation types between entities on non-stationary data with a weak supervision.
\bibliography{reference}
\end{document}